\documentclass[superscriptaddress,twocolumn,showpacs,
amssymb,amsmath,nobibnotes,aps,prd,longbibliography,
nofootinbib]{revtex4-2}
\pdfoutput=1
\usepackage{graphicx,subfig,bm,color,psfrag,hyperref}
\usepackage{amsfonts}
\usepackage{lipsum}
\usepackage{mathtools}
\usepackage{verbatim}
\usepackage[normalem]{ulem}
\usepackage[dvipsnames]{xcolor}
\usepackage{booktabs}
\hypersetup{colorlinks,linkcolor={blue},citecolor={red},urlcolor={cyan}}  

\begin{document}

\title{When Dark Energy Turns On: Constraints on a Critical Emergence Model}

\author{Mahdi Najafi}
\email{mahdinajafi12676@yahoo.com}
\email{mahdi.najafi@uniroma1.it}
\affiliation{Dipartimento di Fisica, Univ. La Sapienza, P. le A. Moro 2, Roma, Italy}
\affiliation{Department of Physics, K.N. Toosi University of Technology, P.O. Box 15875-4416, Tehran, Iran}
\affiliation{PDAT Laboratory, Department of Physics, K.N. Toosi University of Technology, P.O. Box 15875-4416, Tehran, Iran}

\author{Mahdi Habibollahi}
\email{mahdii1379hb@gmail.com}
\affiliation{Department of Physics, K.N. Toosi University of Technology, P.O. Box 15875-4416, Tehran, Iran}
\affiliation{PDAT Laboratory, Department of Physics, K.N. Toosi University of Technology, P.O. Box 15875-4416, Tehran, Iran}

\author{Masoume Reyhani}
\email{masoume.r1079@gmail.com}
\affiliation{Department of Astronomy and High Energy Physics, Faculty of Physics,
Kharazmi University, Tehran, Iran.}
\affiliation{PDAT Laboratory, Department of Physics, K.N. Toosi University of Technology, P.O. Box 15875-4416, Tehran, Iran}

\author{Eleonora Di Valentino}
\email{e.divalentino@sheffield.ac.uk}
\affiliation{School of Mathematical and Physical Sciences, University of Sheffield, Hounsfield Road, Sheffield S3 7RH, United Kingdom}

\author{Supriya Pan}
\email{supriya.maths@presiuniv.ac.in}
\affiliation{Department of Mathematics, Presidency University, 86/1 College Street, Kolkata 700073, India}
\affiliation{Institute of Systems Science, Durban University of Technology, Durban 4000, Republic of South Africa}

\author{Javad T. Firouzjaee}
\email{firouzjaee@kntu.ac.ir}
\affiliation{Department of Physics, K.N. Toosi University of Technology, P.O. Box 15875-4416, Tehran, Iran}
\affiliation{PDAT Laboratory, Department of Physics, K.N. Toosi University of Technology, P.O. Box 15875-4416, Tehran, Iran}
\affiliation{School of Physics, Institute for Research in Fundamental Sciences (IPM), P.O. Box 19395-5531, Tehran, Iran}

\author{Weiqiang Yang}
\email{d11102004@163.com}
\affiliation{Department of Physics, Liaoning Normal University, Dalian, 116029, People's Republic of China}

\begin{abstract}
We investigate a specific emergent dark energy scenario, known as critically emergent dark energy (CEDE), in which dark energy is effectively absent in the early Universe and becomes dynamically relevant only after a critical cosmic epoch through a phase transition. We constrain this model using recent cosmological observations, including cosmic microwave background (CMB) data from \emph{Planck} 2018, baryon acoustic oscillation (BAO) measurements from SDSS and DESI DR2, and two independent Type Ia supernova compilations, PantheonPlus and Union3. 
Our results show that within the CEDE framework a dark energy phase transition is not ruled out. In particular, CMB-only, CMB+SDSS, and CMB+DESI datasets provide evidence for a nonzero transition scale factor and, according to standard statistical indicators such as $\Delta\chi^2$ and Bayesian evidence, can favor CEDE over the $\Lambda$CDM model. At the same time, we find that CEDE does not fully resolve the Hubble constant tension. 
Overall, our analysis indicates that dark energy models featuring a phase transition remain a viable and phenomenologically interesting extension of the standard cosmological framework. Upcoming high-precision cosmological surveys will be essential to further assess whether such emergent dark energy scenarios represent a genuine departure from $\Lambda$CDM or an effective description of current data.
\end{abstract}

\maketitle
\section{Introduction}
\label{sec-introduction}

The standard cosmological model, $\Lambda$-Cold Dark Matter ($\Lambda$CDM), provides a remarkably successful and economical description of the Universe, accurately fitting a wide range of high-precision cosmological observations. Within the framework of Einstein's General Theory of Relativity (GR), $\Lambda$CDM assumes the presence of a cosmological constant $\Lambda$, interpreted as dark energy (DE), together with cold dark matter (CDM) and ordinary matter. Despite its empirical success, the physical nature of both DE and CDM remains unknown, and their apparent independent conservation lacks a fundamental explanation. Moreover, $\Lambda$CDM faces long-standing conceptual challenges, most notably the cosmological constant problem~\cite{Weinberg:1988cp} and the cosmic coincidence problem~\cite{Zlatev:1998tr}.

Beyond these theoretical issues, recent years have seen the emergence of significant observational tensions that challenge the internal consistency of the $\Lambda$CDM framework. Among them, the most prominent is the Hubble constant tension~\cite{Verde:2019ivm,DiValentino:2020zio,DiValentino:2021izs,Perivolaropoulos:2021jda,Schoneberg:2021qvd,Shah:2021onj,Abdalla:2022yfr,DiValentino:2022fjm,Kamionkowski:2022pkx,Giare:2023xoc,Hu:2023jqc,Verde:2023lmm,DiValentino:2024yew,CosmoVerseNetwork:2025alb,Ong:2025cwv}, namely the statistically significant discrepancy between the value of the present-day expansion rate inferred from early-Universe probes and that obtained from direct late-Universe measurements. This discrepancy first exceeded the $5\sigma$ level when comparing the Hubble constant inferred from \emph{Planck} Cosmic Microwave Background (CMB) observations, interpreted within the $\Lambda$CDM framework~\cite{Planck:2018vyg}, with local distance-ladder measurements from the SH0ES collaboration~\cite{Riess:2021jrx}.
More recently, the tension has further increased. By combining CMB measurements from \emph{Planck}, \emph{SPT-3G}, and \emph{ACT}~\cite{Planck:2018vyg,Planck:2018nkj,ACT:2020gnv,ACT:2025fju,SPT-3G:2025bzu}, and comparing them with a homogeneous compilation of direct late-Universe determinations assembled by the H$_0$ Distance Network (H0DN), the discrepancy has reached approximately $7.1\sigma$~\cite{H0DN:2025lyy}, representing the most robust manifestation of the Hubble constant tension to date.
The persistence of this early-late discrepancy is particularly intriguing because the late-Universe determinations rely on a broad range of independent observational techniques and astrophysical systems, yet consistently converge on higher values of $H_0$~\cite{Freedman:2020dne,Birrer:2020tax,Riess:2021jrx,Anderson:2023aga,Scolnic:2023mrv,Jones:2022mvo,Anand:2021sum,Freedman:2021ahq,Uddin:2023iob,Huang:2023frr,Li:2024yoe,Pesce:2020xfe,Kourkchi:2020iyz,Schombert:2020pxm,Blakeslee:2021rqi,deJaeger:2022lit,Murakami:2023xuy,Breuval:2024lsv,Freedman:2024eph,Riess:2024vfa,Vogl:2024bum,Scolnic:2024hbh,Said:2024pwm,Boubel:2024cqw,Scolnic:2024oth,Li:2025ife,Jensen:2025aai,Riess:2025chq,Benisty:2025tct,Newman:2025gwg,Stiskalek:2025ktq,H0DN:2025lyy,Agrawal:2025tuv,Bhardwaj:2025kbw}. Reconciling this discrepancy within a single cosmological framework has proven extremely challenging and has motivated extensive exploration of physics beyond $\Lambda$CDM.
Proposed explanations span a wide range of physical mechanisms. These include extensions of the standard cosmological parameter space~\cite{DiValentino:2016hlg}, modifications of gravity~\cite{Koyama:2015vza,Cai:2015emx,Nojiri:2017ncd,DiValentino:2015bja,Zumalacarregui:2020cjh,Odintsov:2020qzd,Adi:2020qqf,DeFelice:2020cpt,Pogosian:2021mcs,CANTATA:2021asi,Bahamonde:2021gfp,Schiavone:2022wvq,Specogna:2023nkq,Tiwari:2023jle,Hogas:2023pjz,Wen:2023wes,Pitrou:2023swx,Montani:2024pou,Dwivedi:2024okk,Ishak:2024jhs,Specogna:2024euz,AtacamaCosmologyTelescope:2025nti,Giare:2025ath,Akarsu:2024qsi,Akarsu:2024nas,Hogas:2025ahb,Jalali:2025cbi}, vacuum phase transitions~\cite{DiValentino:2017rcr}, evolving or late-time dark energy scenarios~\cite{Dutta:2018vmq,vonMarttens:2019ixw,Akarsu:2019hmw,DiValentino:2020naf,DiValentino:2020vnx,Yang:2021flj,DiValentino:2021rjj,Heisenberg:2022lob,Giare:2023xoc,Adil:2023exv,Gomez-Valent:2023uof,Lapi:2023plb,Krolewski:2024jwj,Bousis:2024rnb,Tang:2024gtq,Jiang:2024xnu,Manoharan:2024thb,Specogna:2025guo,Ozulker:2025ehg,Lee:2025pzo}, early-Universe solutions such as early dark energy~\cite{Poulin:2018cxd,Smith:2019ihp,Niedermann:2019olb,Krishnan:2020obg,Schoneberg:2021qvd,Ye:2021iwa,Poulin:2021bjr,Niedermann:2021vgd,deSouza:2023sqp,Poulin:2023lkg,Cruz:2023lmn,Niedermann:2023ssr,Vagnozzi:2023nrq,Efstathiou:2023fbn,Simon:2024jmu,Cervantes-Cota:2023wet,Garny:2024ums,Giare:2024akf,Giare:2024syw,Poulin:2024ken,Pedrotti:2024kpn,Kochappan:2024jyf,Poulin:2025nfb,Smith:2025zsg}, late-time or local transitions in the expansion history~\cite{DiValentino:2019exe,Alestas:2021luu,Ruchika:2023ugh,Frion:2023xwq,Ruchika:2024ymt}, dark energy models with negative energy density or sign-switching behavior~\cite{Akarsu:2019hmw,Visinelli:2019qqu,Ye:2020btb,Calderon:2020hoc,Akarsu:2021fol,Sen:2021wld,DiGennaro:2022ykp,Akarsu:2022typ,Ong:2022wrs,Akarsu:2023mfb,Anchordoqui:2023woo,Adil:2023exv,Akarsu:2024qsi,Halder:2024uao,Anchordoqui:2024gfa,Akarsu:2024eoo,Yadav:2024duq,Paraskevas:2024ytz,Gomez-Valent:2024tdb,Toda:2024ncp,Gomez-Valent:2024ejh,Akarsu:2025gwi,Souza:2024qwd,Soriano:2025gxd,Akarsu:2025ijk,Escamilla:2025imi,Bouhmadi-Lopez:2025ggl,Ghafari:2025eql}, emergent dark energy scenarios~\cite{Pan:2019hac,Yang:2020zuk,Yang:2021eud}, solutions involving dark matter or neutrino physics~\cite{DiValentino:2017oaw,Anchordoqui:2022gmw,Pan:2023frx,Allali:2024anb,Co:2024oek,Aboubrahim:2024spa}, and modifications to the physics of recombination~\cite{Hart:2017ndk,Hart:2019dxi,Sekiguchi:2020teg,Hart:2021kad,Lee:2022gzh,Chluba:2023xqj,Greene:2023cro,Greene:2024qis,Baryakhtar:2024rky,Seto:2024cgo,Mirpoorian:2024fka,Lynch:2024hzh,Toda:2024ncp,Schoneberg:2024ynd,Smith:2025uaq,GarciaEscudero:2025lef,Toda:2025kcq}. Despite their diversity, these approaches share the common goal of reconciling early- and late-Universe determinations of the Hubble constant within a consistent cosmological framework. 

In parallel with the Hubble constant tension, recent large-scale structure measurements have provided independent evidence pointing toward departures from a cosmological constant. In particular, the Dark Energy Spectroscopic Instrument (DESI)~\cite{DESI:2024mwx,DESI:2025zgx} has reported a preference for dynamical dark energy (DDE) when Baryon Acoustic Oscillation (BAO) measurements are combined with Type Ia supernova (SNIa) data. Assuming the Chevallier-Polarski-Linder (CPL)~\cite{Chevallier:2000qy,Linder:2002et} parametrization of the dark energy equation of state, this preference reaches a statistical significance of approximately $3.2$-$3.4\sigma$~\cite{DES:2025sig,Hoyt:2026fve} and remains stable across successive SNIa data reanalysis.
Although this evidence is intrinsically model dependent and parametrization driven, it has been widely explored in the literature using different combinations of cosmological datasets and alternative dark energy parametrizations~\cite{DESI:2025fii,DESI:2024mwx,DESI:2025zgx,Cortes:2024lgw,Shlivko:2024llw,Luongo:2024fww,Gialamas:2024lyw,Dinda:2024kjf,Najafi:2024qzm,Wang:2024dka,Ye:2024ywg,Tada:2024znt,Carloni:2024zpl,Chan-GyungPark:2024mlx,Ferri:2025tuo,DESI:2024kob,Bhattacharya:2024hep,Ramadan:2024kmn,Pourojaghi:2024tmw,Giare:2024gpk,Reboucas:2024smm,Giare:2024ocw,Chan-GyungPark:2024brx,Li:2024qus,Jiang:2024xnu,RoyChoudhury:2024wri,Li:2025cxn,Wolf:2025jlc,Shajib:2025tpd,Giare:2025pzu,Chaussidon:2025npr,Kessler:2025kju,Pang:2025lvh,Roy:2024kni,RoyChoudhury:2025dhe,Paliathanasis:2025cuc,Scherer:2025esj,Giare:2024oil,Liu:2025mub,Teixeira:2025czm,Santos:2025wiv,Specogna:2025guo,Reyhani:2024cnr,Sabogal:2025jbo,Cheng:2025lod,Herold:2025hkb,Cheng:2025hug,Ozulker:2025ehg,Lee:2025pzo,Ormondroyd:2025iaf,Silva:2025twg,Ishak:2025cay,Fazzari:2025lzd,Smith:2025icl,Zhang:2025lam,Cheng:2025yue}. Taken together, these results suggest that the assumption of a strictly constant dark energy density may be overly restrictive and motivate further investigation of scenarios in which DE evolves dynamically at late times.

In this context, models in which DE emerges dynamically at late times, potentially through a phase transition, provide a particularly compelling phenomenological avenue to investigate. While the literature exploring extensions of $\Lambda$CDM is extensive, no single scenario has so far been able to simultaneously accommodate all available cosmological observations. As a result, alternative cosmological models remain well motivated as long as they are not ruled out at a statistically significant level.
In this work, we focus on a DE model endowed with a late-time phase transition, known as critically emergent dark energy (CEDE). This scenario is motivated by the Ginzburg-Landau theory of phase transitions~\cite{Ginzburg:1950sr} and was originally proposed as a possible mechanism to alleviate the Hubble constant tension~\cite{Banihashemi:2018has,Banihashemi:2018oxo,Banihashemi:2020wtb}. In the CEDE framework, DE has no effective presence at early times and emerges only after a critical redshift, corresponding to a cosmological phase transition in the DE sector. The physical appeal of this framework lies in the fact that cosmic acceleration is a relatively recent phenomenon, making it natural to consider models in which DE becomes dynamically relevant only at late epochs.
Beyond its original motivation in the context of the $H_0$ tension, CEDE also provides a natural phenomenological framework to explore recent indications of DDE emerging from large-scale structure observations. In particular, a late-time phase transition offers an alternative to smooth equation-of-state parametrizations, allowing for a more abrupt departure from a cosmological constant-like behavior that may be compatible with current data.
Motivated by these considerations, we constrain the CEDE model using a comprehensive set of cosmological observations, including CMB measurements, BAO data from both the Sloan Digital Sky Survey (SDSS)~\cite{eBOSS:2020yzd} and DESI, and SNIa from the PantheonPlus~\cite{Scolnic:2021amr} and Union3~\cite{Rubin:2023ovl} compilations. In addition, we perform a detailed model-comparison analysis to assess the statistical viability of CEDE relative to the standard $\Lambda$CDM paradigm.

The manuscript is organized as follows. In Section~\ref{sec-2} we introduce the CEDE model and discuss its evolution at both the background and perturbative levels. Section~\ref{sec-data} describes the observational datasets and the numerical methodology adopted in our analysis. The results are presented and discussed in Section~\ref{sec-results}. Finally, in Section~\ref{sec-summary} we summarize the main findings and present our conclusions.

\section{Critically Emergent Dark Energy}
\label{sec-2}

We assume that, on sufficiently large scales, the Universe can be described as homogeneous and isotropic and is therefore well approximated by a Friedmann-Lema\^{i}tre-Robertson-Walker (FLRW) spacetime. The corresponding line element is
\begin{equation}
ds^2 = -dt^2 + a^2(t)\,\delta_{ij}\,dx^i dx^j ,
\end{equation}
where $a(t)$ denotes the scale factor. Assuming spatial flatness, the expansion history of the Universe is governed by the Hubble function
\begin{eqnarray}\label{Hubble-expansion}
\frac{H}{H_0} = \left[\Omega_{m0}(1+z)^3 + \Omega_{r0}(1+z)^4 + \frac{\rho_{\nu}}{\rho_{c0}} + \widetilde{\Omega}_{\rm DE}(z) \right]^{1/2}.
\end{eqnarray}
Here we assume that gravity is described by General Relativity, that matter is minimally coupled to the gravitational sector, and that the individual cosmic fluids do not interact with each other. In Eq.~(\ref{Hubble-expansion}), $\Omega_{m0}$ and $\Omega_{r0}$ denote the present-day density parameters of matter and radiation, respectively, $\rho_{\nu}$ is the energy density of the neutrino sector, $\rho_{c0}$ is the critical energy density today (at $a=1$), and $\widetilde{\Omega}_{\rm DE}(z)\equiv\rho_{\rm DE}(z)/\rho_{c0}$ denotes the dimensionless dark energy density.
Solving the conservation equation for dark energy,
\begin{equation}
\dot{\rho}_{\rm DE}+3H(1+w_{\rm DE})\rho_{\rm DE}=0,
\end{equation}
where the dot denotes differentiation with respect to cosmic time and $w_{\rm DE}$ is the dark energy equation-of-state (EoS) parameter, leads to the general expression
\begin{eqnarray}
\widetilde{\Omega}_{\rm DE}(z)=\Omega_{\rm DE,0}\exp\!\left[-3\int_{0}^{z}\frac{1+w_{\rm DE}(z')}{1+z'}\,dz'\right].
\end{eqnarray}
This expression can be evaluated once a specific functional form for $w_{\rm DE}(z)$ is prescribed.
An alternative and often convenient approach is to parametrize directly the dark energy density, rather than its equation of state. In this work, we adopt the critically emergent dark energy (CEDE) parametrization proposed in Ref.~\cite{Banihashemi:2020wtb},
\begin{eqnarray}\label{model-cede}
\widetilde{\Omega}_{\rm DE}(z)=\Omega_{\rm DE,0}\left(\frac{z_c-z}{z_c}\right)^{1/2},
\end{eqnarray}
where $z_c$ denotes the critical redshift associated with the phase transition.
Since this parametrization is physically meaningful only for $z<z_c$, we explicitly define the model as
\begin{eqnarray}\label{model-redefine}
\widetilde{\Omega}_{\rm DE}(z)=
\begin{cases}
\Omega_{\rm DE,0}\left(\dfrac{z_c-z}{z_c}\right)^{1/2}, & z<z_c,\\[2mm]
0, & z\ge z_c,
\end{cases}
\end{eqnarray}
thereby enforcing the absence of dark energy at earlier times. This prescription is consistent with the emergent nature of the CEDE scenario, in which dark energy becomes dynamically relevant only after the phase transition.
From the dark energy conservation equation, the effective equation of state of dark energy can be written as
\begin{eqnarray}\label{eq:wDE}
w_{\rm DE}(z) = -1 + \frac{1+z}{3\,\widetilde{\Omega}_{\rm DE}(z)} 
\frac{d}{dz}\widetilde{\Omega}_{\rm DE}(z).
\end{eqnarray}
For the CEDE parametrization defined in Eq.~(\ref{model-redefine}), this yields
\begin{eqnarray}\label{eq:wDE_cede}
w_{\rm DE}(z) = -1 - \frac{1+z}{6\,(z_c - z)},
\end{eqnarray}
which is valid in the regime $z<z_c$.\footnote{For $z>z_c$, the dark energy density vanishes by construction, and therefore the equation of state is defined only in the region $z<z_c$.}
From this expression, it follows that after the phase transition at $z=z_c$, the DE component lies in the phantom regime, $w_{\rm DE}<-1$, throughout its evolution. This behavior is a direct consequence of the adopted density parametrization.\footnote{Recent DESI analyses~\cite{DESI:2025zgx} have reported a preference for quintessence-like behavior at the present epoch when assuming a CPL parametrization~\cite{Chevallier:2000qy,Linder:2002et}. Since CEDE is defined through a density parametrization associated with a phase transition, rather than a smooth equation-of-state ansatz, such results are not directly applicable. More general DE frameworks allowing transitions or crossings of the phantom divide have been widely discussed in the literature, e.g. in the context of quintom models~\cite{Cai:2009zp}.}

\begin{figure}
    \centering
\includegraphics[width=0.5\textwidth]{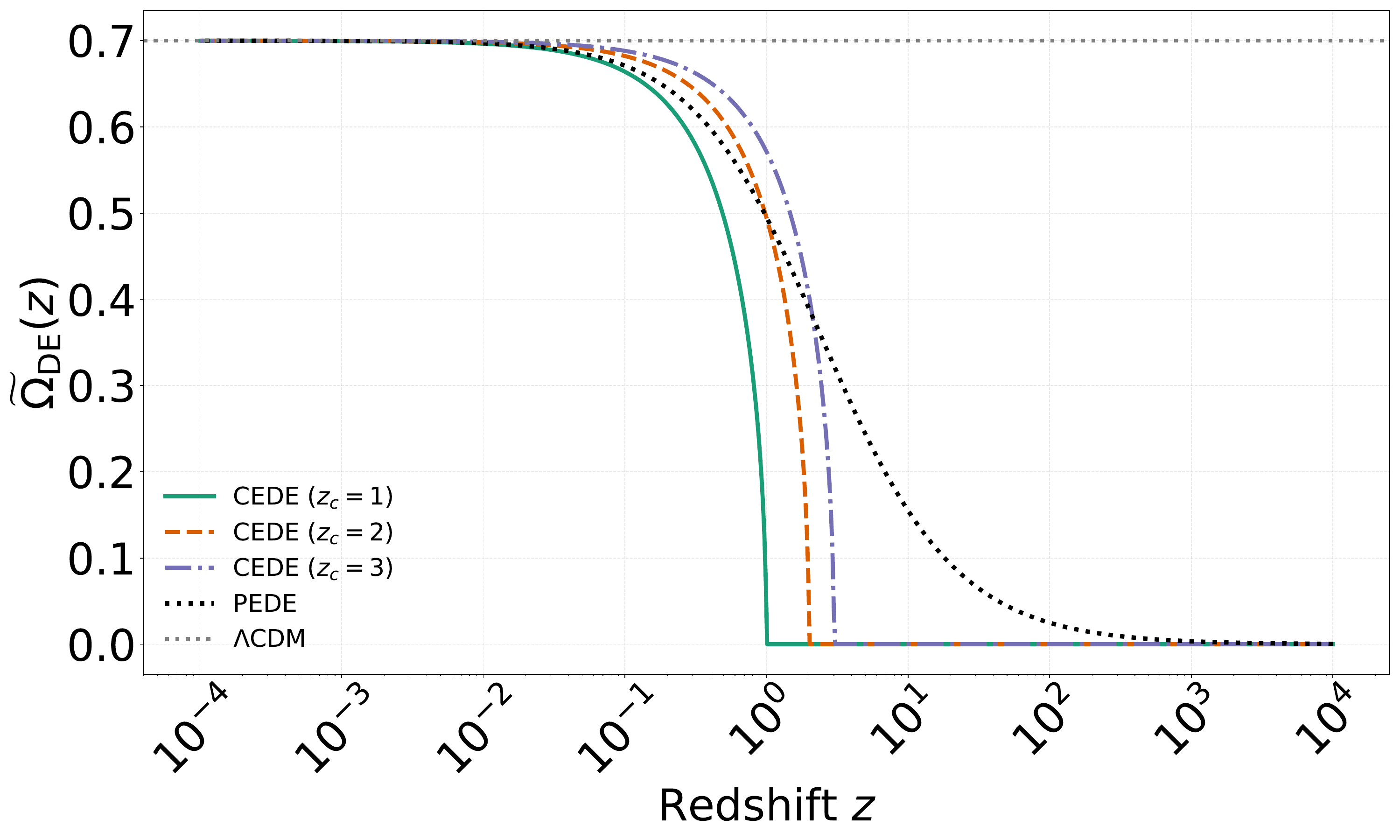}
\caption{Redshift evolution of the dimensionless DE density $\widetilde{\Omega}_{\rm DE}(z)$ in the CEDE model for different values of the critical redshift $z_c$, compared with the PEDE and $\Lambda$CDM models. In all cases, we fix the present-day DE density to $\Omega_{\rm DE,0}=0.7$.}
    \label{fig:OmegaDE}
\end{figure}

In Fig.~\ref{fig:OmegaDE} we show the redshift evolution of the DE density $\widetilde{\Omega}_{\rm DE}(z)$, as defined in Eq.~(\ref{model-redefine}), for several representative values of the critical redshift $z_c$. For comparison, we also include another emergent DE scenario, namely the phenomenologically emergent dark energy (PEDE)~\cite{Li:2019yem,Pan:2019hac},\footnote{In the PEDE model, the dark energy density evolves as $\widetilde{\Omega}_{\rm DE}(z)=\Omega_{\rm DE,0}\left[1-\tanh\!\left(\log_{10}(1+z)\right)\right]$~\cite{Li:2019yem,Pan:2019hac}.} as well as the standard $\Lambda$CDM model.
The sharp onset of $\widetilde{\Omega}_{\rm DE}$ from a vanishing value clearly distinguishes the CEDE scenario from PEDE, while both models asymptotically approach a constant dark energy density at late times.

We now investigate the impact of the CEDE model at the level of linear cosmological perturbations, which is essential for assessing its implications for the formation of large-scale structure. Since dark energy emerges only below a critical redshift, as defined in Eq.~(\ref{model-redefine}), the perturbative effects associated with CEDE are expected to become relevant only in this late-time regime.
To study linear perturbations, we consider scalar perturbations around the homogeneous and isotropic background in a specific gauge. In this work, we adopt the synchronous gauge, for which the perturbed spatially flat FLRW line element is given by~\cite{Ma:1995ey}
\begin{equation}
ds^2 = a^2(\tau)\left[-d\tau^2 + (\delta_{ij}+h_{ij})\,dx^i dx^j\right],
\end{equation}
where $\tau$ denotes conformal time, $\delta_{ij}$ is the unperturbed spatial metric, and $h_{ij}$ represents the metric perturbations. The scale factor is expressed as a function of conformal time throughout. Using this metric, the perturbed Einstein equations can be derived in the standard way. 
Following Ref.~\cite{Ma:1995ey}, we introduce the dimensionless density contrast $\delta_i\equiv\delta\rho_i/\rho_i$ for the $i$-th fluid and the velocity divergence $\theta_i\equiv i\,\kappa^j v_j$. In Fourier space, the linear evolution equations for each fluid component can be written as
\begin{eqnarray}
\delta'_i &=& - (1+w_i)\left(\theta_i + \frac{h'}{2}\right)
-3\mathcal{H}\left(\frac{\delta P_i}{\delta\rho_i}-w_i\right)\delta_i \nonumber\\
&& -9\mathcal{H}^2\left(\frac{\delta P_i}{\delta\rho_i}-c^2_{a,i}\right)
(1+w_i)\frac{\theta_i}{\kappa^2}, \label{per1} \\
\theta'_i &=& -\mathcal{H}\left(1-3\frac{\delta P_i}{\delta\rho_i}\right)\theta_i
+ \frac{\delta P_i/\delta\rho_i}{1+w_i}\,\kappa^2\,\delta_i
-\kappa^2\sigma_i. \label{per2}
\end{eqnarray}
Here a prime denotes differentiation with respect to conformal time, $\mathcal{H}$ is the conformal Hubble parameter, $h$ is the synchronous-gauge metric perturbation, and $\kappa$ is the comoving wavenumber. The quantity $\sigma_i$ denotes the anisotropic stress of the $i$-th fluid; although it is kept in the general equations, we neglect anisotropic stress in the numerical analysis and set $\sigma_i=0$.
The quantity $\delta P_i/\delta\rho_i$ corresponds to the rest-frame sound speed squared of the $i$-th fluid (for dark energy, $\delta P_{\rm DE}/\delta\rho_{\rm DE}\equiv c^2_{\rm s,DE}$), while
\begin{equation}
c^2_{a,i}=w_i-\frac{w_i'}{3\mathcal{H}(1+w_i)}
\end{equation}
is the adiabatic sound speed. In our analysis, we fix the dark energy sound speed to $c^2_{\rm s,DE}=1$, as commonly assumed for minimally coupled scalar-field models, and set the matter sound speed to $c^2_{\rm s,m}=0$. These perturbation equations are implemented consistently in the statistical analysis of the CEDE model. 

We conclude this section by considering the growth of matter perturbations in the presence of emergent dark energy, which provides insight into how the CEDE scenario affects the formation of large-scale structure. At the linear level, the evolution of the matter density contrast, $\delta_m\equiv\delta\rho_m/\rho_m$, is governed by~\cite{Abramo:2007iu}
\begin{eqnarray}\label{growth-equation}
\ddot{\delta}_{\rm m}+2H\dot{\delta}_{\rm m}-\frac{3H^2}{2}\Omega_{\rm m}(t)\,\delta_{\rm m}=0.
\label{matter}
\end{eqnarray}

In this equation, the impact of CEDE enters through the background expansion rate encoded in the Hubble function $H(t)$. Consequently, modifications to the expansion history induced by emergent dark energy directly influence the growth of matter perturbations. By analyzing the evolution of $\delta_m$, we can therefore quantify the effect of the CEDE model on structure formation.

We compute the quantity $f\sigma_8(a)$, defined as~\cite{Nesseris:2007pa}
\begin{align}
f\sigma_8(a) \equiv f(a)\,\sigma_8(a),
\end{align}
where $f(a)=d\ln\delta_m/d\ln a$ is the linear growth rate of matter perturbations and quantifies the formation and evolution of cosmic structures. The quantity $\sigma_8(a)$ denotes the root-mean-square fluctuation of the linear density field within a sphere of radius $R=8\,h^{-1}\mathrm{Mpc}$, and is given by
\begin{equation}
\sigma_8(a)=\sigma_{8,0}\,\frac{\delta_m(a)}{\delta_m(1)}.
\end{equation}

\begin{figure}
    \centering
    \includegraphics[width=1.05\linewidth]{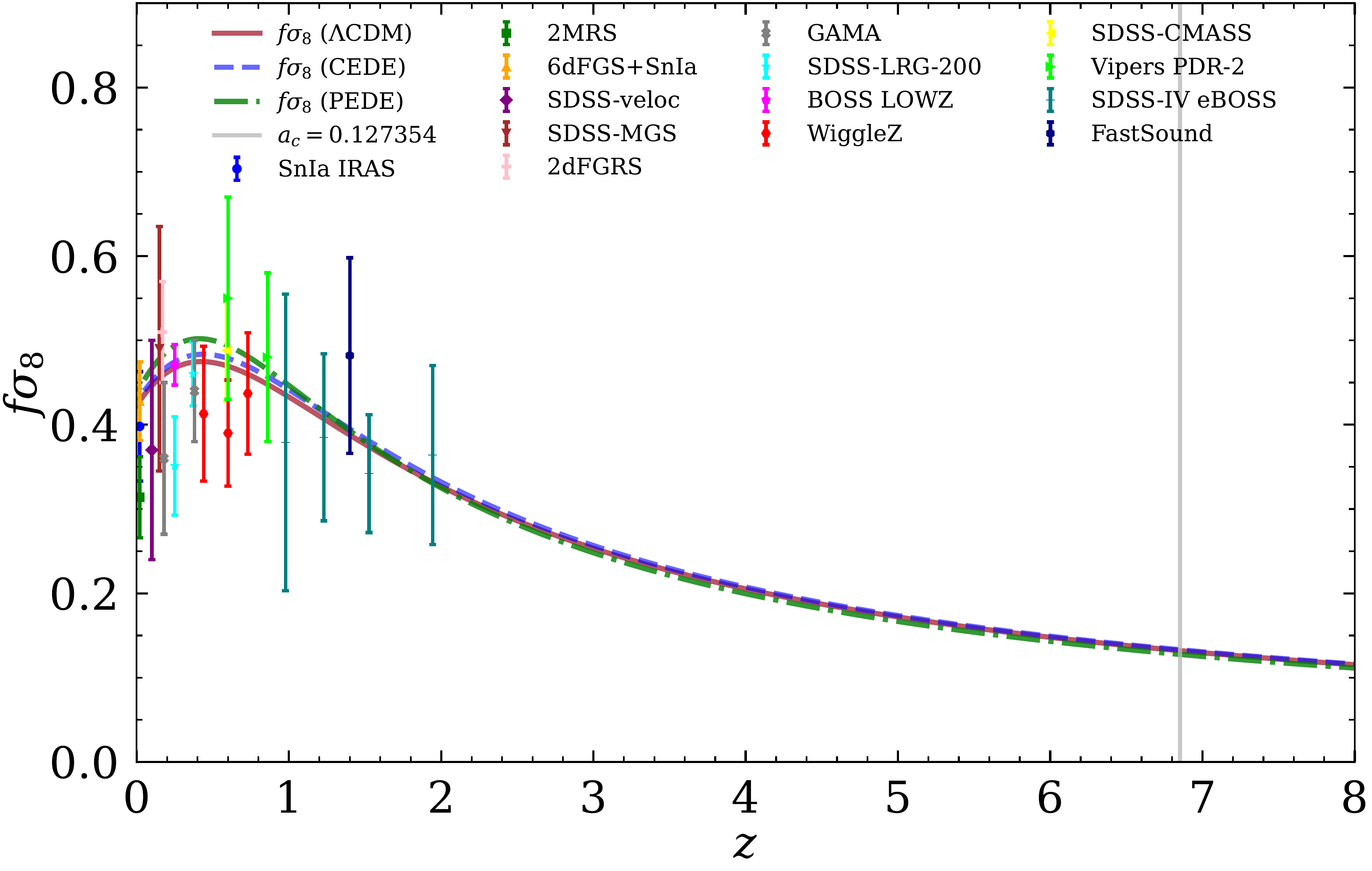}
    \caption{Evolution of the growth-rate observable $f\sigma_8(z)$ for the CEDE model, compared with the PEDE and $\Lambda$CDM scenarios. The curves are shown alongside observational measurements from various large-scale structure surveys. The vertical grey line indicates the redshift at which the dark energy equation of state $w_{\rm DE}$ diverges in the CEDE model.}
    \label{fig:fsigma8}
\end{figure}

To illustrate the evolution of the growth-rate observable $f\sigma_8(z)$, we compute the normalized matter density contrast $\delta_m(a)/\delta_m(a=1)$ by numerically solving Eq.~(\ref{growth-equation}).\footnote{\url{https://github.com/camarman/MDP-Ls.git}~\cite{Akarsu:2025ijk}} We adopt the initial conditions $a_{\rm ini}=10^{-3}$, $\delta_m(a_{\rm ini})=0.002$, and $\delta_m'(a_{\rm ini})=2.0$, corresponding to the growing mode in the early Universe. The calculations are performed using the best-fit cosmological parameters obtained from the CMB+SDSS dataset, namely $\sigma_{8,0}^{\rm CEDE}=0.82548$, $\sigma_{8,0}^{\Lambda\rm CDM}=0.814011$, $\Omega_{m,0}^{\rm CEDE}=0.302963$, $\Omega_{m,0}^{\Lambda\rm CDM}=0.306859$, and $a_c=0.127354$.
Figure~\ref{fig:fsigma8} shows the theoretical evolution of $f\sigma_8(z)$ predicted by the CEDE model, together with the corresponding predictions from the PEDE~\cite{Li:2019yem,Pan:2019hac} and $\Lambda$CDM scenarios, compared with observational measurements at different redshifts summarized in Table~\ref{tab:fsigma8}. We find that the evolution of $f\sigma_8(z)$ in the CEDE framework closely tracks that of $\Lambda$CDM over the redshift range considered. Relative to the PEDE model, CEDE displays a broadly similar behavior, with only mild deviations appearing at low redshift.

\begin{table}
\centering
 \resizebox{0.8\columnwidth}{!}{
\small
\renewcommand{\arraystretch}{0.7}
\begin{tabular}{l c c c} 
\hline 
\multicolumn{4}{c}{\vspace{0.01mm}} \\
\textbf{Dataset} & \textbf{$z$} & $\mathbf{f\sigma_8(z)}$ & \textbf{Ref.} \\
\multicolumn{4}{c}{\vspace{0.01cm}} \\
\hline\hline 
\multicolumn{4}{c}{\vspace{0.01mm}} \\
~2MTF~ & $0.001$ & $0.505 \pm 0.085$ & \cite{Howlett:2017asq} \\
~6dFGS+SNIa~ & $0.020$ & $0.4280 \pm 0.0465$ & \cite{Huterer:2016uyq}\\
~IRAS+SNIa~ & $0.020$ & $0.398 \pm 0.065$ & \cite{Hudson:2012gt, Turnbull:2011ty}\\
~2MASS~ & $0.020$ & $0.314 \pm 0.048$ & \cite{Hudson:2012gt, Davis:2010sw}\\
~2dFGRS~ & $0.170$ & $0.510 \pm 0.060$ & \cite{Song:2008qt}\\
~GAMA~ & $0.180$ & $0.360 \pm 0.090$ & \cite{Blake:2013nif}\\
~GAMA~ & $0.380$ & $0.440 \pm 0.060$ & \cite{Blake:2013nif}\\
~SDSS-IV (MGS)~ & $0.150$ & $0.530 \pm 0.160$ & \cite{eBOSS:2020yzd} \\
~SDSS-IV (BOSS Galaxy)~ & $0.380$ & $0.497 \pm 0.045$ & \cite{eBOSS:2020yzd}\\
~SDSS-IV (BOSS Galaxy)~ & $0.510$ & $0.459 \pm 0.038$ & \cite{eBOSS:2020yzd}\\
~SDSS-IV (eBOSS LRG)~ & $0.700$ & $0.473 \pm 0.041$ & \cite{eBOSS:2020yzd}\\
~SDSS-IV (eBOSS ELG)~ & $0.850$ & $0.315 \pm 0.095$ & \cite{eBOSS:2020yzd}\\
~WiggleZ~ & $0.44$ & $0.413 \pm 0.080$ & \cite{Blake:2012pj} \\
~WiggleZ~ & $0.600$ & $0.390 \pm 0.063$ & \cite{Blake:2012pj}\\
~WiggleZ~ & $0.730$ & $0.437 \pm 0.072$ & \cite{Blake:2012pj}\\
~Vipers PDR-2~ & $0.600$ & $0.550 \pm 0.120$ & \cite{Pezzotta:2016gbo} \\
~Vipers PDR-2~ & $0.860$ & $0.400 \pm 0.110$ & \cite{Pezzotta:2016gbo} \\
~FastSound~ & $1.400$ & $0.482 \pm 0.116$ & \cite{Okumura:2015lvp} \\
~SDSS-IV (eBOSS Quasar)~ & $0.978$ & $0.379 \pm 0.176$ & \cite{eBOSS:2018yfg}\\
~SDSS-IV (eBOSS Quasar)~ & $1.230$ & $0.385 \pm 0.099$ & \cite{eBOSS:2018yfg}\\
~SDSS-IV (eBOSS Quasar)~ & $1.526$ & $0.342 \pm 0.070$ & \cite{eBOSS:2018yfg}\\
~SDSS-IV (eBOSS Quasar)~ & $1.944$ & $0.364 \pm 0.106$ & \cite{eBOSS:2018yfg}\\ 
\hline

\end{tabular}}
\caption{Compilation of observational measurements of $f\sigma_8$ at different redshifts used in this work. The table reports the effective redshift, the measured value of $f\sigma_8$, and the corresponding reference for each dataset.}
\label{tab:fsigma8}
 \label{tab:fsigma8}
\end{table}

\begin{table}
\centering
\renewcommand{\arraystretch}{1.5}
\begin{tabular}{l @{\hspace{2cm}} c}
\hline 
\textbf{Parameter} & \textbf{Prior} \\
\hline \hline
$\Omega_\mathrm{b} h^2$ & $[0.005, 0.1]$ \\
$\Omega_\mathrm{c} h^2$ & $[0.01, 0.99]$ \\
$\log(10^{10} A_\mathrm{s})$ & $[1.61, 3.91]$ \\
$n_\mathrm{s}$ & $[0.8, 1.2]$ \\
$\tau$ & $[0.01, 0.8]$ \\
$100\theta_\mathrm{MC}$ & $[0.5, 10]$ \\
$a_c$ & $[0.01, 1]$ \\
\hline 
\end{tabular}
\caption{Flat prior ranges adopted for the free cosmological parameters in the analysis.}
\label{tab-priors}
\end{table}

\begin{table}
\centering
\renewcommand{\arraystretch}{1.5}
\begin{tabular}{l @{\hspace{2cm}} c}
\hline 
$\ln \mathcal{B}_{ij}$ & \textbf{interpretation} \\
\hline \hline
$0 \leq | \ln \mathcal{B}_{ij}|  < 1$ & Inconclusive \\
$1 \leq | \ln \mathcal{B}_{ij}|  < 2.5$ & Weak \\
$2.5 \leq | \ln \mathcal{B}_{ij}|  < 5$ & Moderate \\
$5 \leq | \ln \mathcal{B}_{ij}|  < 10$ & Strong \\
$| \ln \mathcal{B}_{ij} | \geq 10$ & Very strong \\

\hline 
\end{tabular}
\caption{Interpretation of the Bayes factor $\ln \mathcal{B}_{ij}$.}
\label{tab-BE}
\end{table}

\section{Statistical methodology and observational datasets}
\label{sec-data}
In this section, we describe the cosmological datasets and the statistical methodology adopted to constrain the model parameters. The CEDE model is characterized by seven free parameters: six standard cosmological parameters of the $\Lambda$CDM model, namely $\Omega_{\rm b}h^2$, $\Omega_{\rm c}h^2$, $\tau_{\rm reio}$, $100\,\theta_{\rm s}$, $\log(10^{10}A_{\rm s})$, and $n_{\rm s}$,\footnote{Here $\Omega_{\rm b}h^2$ is the physical baryon density, $\Omega_{\rm c}h^2$ is the physical cold dark matter density, $\tau_{\rm reio}$ denotes the optical depth to reionization, $\theta_{\rm s}$ is the angular scale of the sound horizon at recombination, $A_{\rm s}$ is the amplitude of the primordial scalar perturbations, and $n_{\rm s}$ is the scalar spectral index.} together with one additional parameter, $a_c$, which denotes the scale factor at the dark energy phase transition.
To explore the parameter space of the model, we modify the publicly available \texttt{CAMB} (Code for Anisotropies in the Microwave Background) code~\cite{Lewis:1999bs,Howlett:2012mh} and perform Markov Chain Monte Carlo (MCMC) analyses using the \texttt{Cobaya} sampler~\cite{Torrado:2020dgo}. The MCMC chains are evolved until convergence is achieved, which is assessed using the Gelman--Rubin diagnostic~\cite{Gelman:1992zz}. For all analyses, we require $R-1<0.03$. The flat prior ranges imposed on the cosmological parameters are summarized in Table~\ref{tab-priors}.  

In addition to parameter constraints, we perform a model-comparison analysis to assess the statistical preference of the CEDE model relative to the standard $\Lambda$CDM scenario. For this purpose, we employ Bayesian evidence.
For a given cosmological model $\mathcal{M}_i$, characterized by a parameter vector $\Theta$, the Bayesian evidence is defined as
\begin{equation}
B_i = \int \mathcal{L}(D|\Theta,\mathcal{M}_i)\,\pi(\Theta|\mathcal{M}_i)\,d\Theta,
\end{equation}
where $\mathcal{L}$ denotes the likelihood of the data $D$ and $\pi$ represents the prior distribution. Taking $\Lambda$CDM as the reference model $\mathcal{M}_j$, we compute the Bayes factor
\begin{equation}
B_{ij}=\frac{B_i}{B_j},
\end{equation}
and report the corresponding relative log-evidence,
\begin{equation}
\ln B_{ij} = \ln B_i - \ln B_j = \ln B_{\rm CEDE} - \ln B_{\Lambda{\rm CDM}}.
\end{equation}
A positive (negative) value of $\ln B_{ij}$ indicates a preference for the CEDE model (for $\Lambda$CDM). The strength of evidence associated with different values of $\ln B_{ij}$ is interpreted using the revised Jeffreys' scale, summarized in Table~\ref{tab-BE}.
For completeness, we also report the difference in the minimum chi-square,
\begin{equation}
\Delta\chi^2 = \chi^2_{\rm min}({\rm CEDE}) - \chi^2_{\rm min}(\Lambda{\rm CDM}),
\end{equation}
such that $\Delta\chi^2<0$ ($>0$) indicates a better (worse) fit of the CEDE model relative to $\Lambda$CDM. 

In the following, we describe the cosmological probes considered in this work:

\begin{enumerate}
    \item \textbf{CMB}: 
    Cosmic microwave background (CMB) observations provide one of the most powerful probes of cosmology and dark energy. In this analysis, we use temperature and polarization anisotropy measurements from the \emph{Planck} 2018 data release (plikTTTEEE+lowl+lowE)~\cite{Planck:2018vyg,Planck:2019nip}, complemented by CMB lensing measurements from ACT~\cite{ACT:2023dou}.

    \item \textbf{BAO}: 
    Baryon acoustic oscillations (BAO) constitute another key cosmological probe. We consider two independent BAO compilations: one from the Sloan Digital Sky Survey (SDSS)~\cite{Beutler:2011hx,Ross:2014qpa,eBOSS:2020yzd} and another from DESI Data Release~2 (DR2)~\cite{DESI:2025zgx}. The DESI DR2 BAO dataset includes cosmic distance measurements from multiple large-scale structure tracers, such as bright galaxies (BGS), emission line galaxies (ELG), luminous red galaxies (LRG), and quasi-stellar objects (QSO). In addition, we include the auto-correlation of Ly-$\alpha$ forest spectra and their cross-correlation with QSOs~\cite{DESI:2025zpo}.

    \item \textbf{SNIa}: 
    Finally, we include Type~Ia supernovae (SNIa), which provided the first direct evidence for the accelerated expansion of the Universe. We consider two independent SNIa compilations: \textbf{PantheonPlus}~\cite{Scolnic:2021amr} and \textbf{Union3}, the latter containing 2,087 supernovae~\cite{Rubin:2023ovl}.
\end{enumerate}

\begin{figure}
    \centering
    \includegraphics[width=0.5\textwidth]{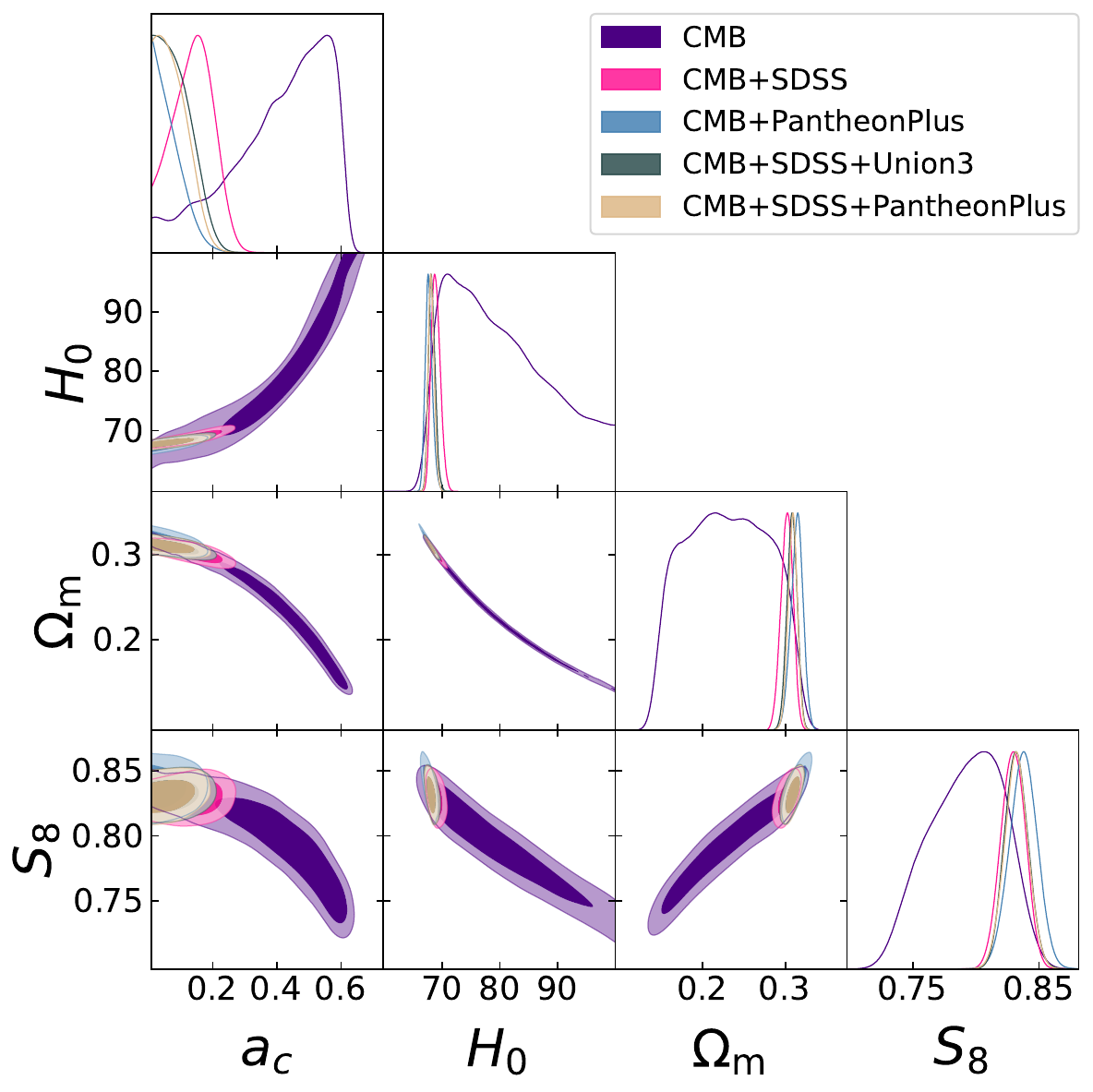}
    \caption{One-dimensional marginalized posterior distributions and two-dimensional joint confidence contours for selected CEDE model parameters, obtained using the combined CMB, SDSS BAO, PantheonPlus, and Union3 datasets.}
    \label{fig:contour1}
\end{figure}
\begin{figure}
    \centering
    \includegraphics[width=0.5\textwidth]{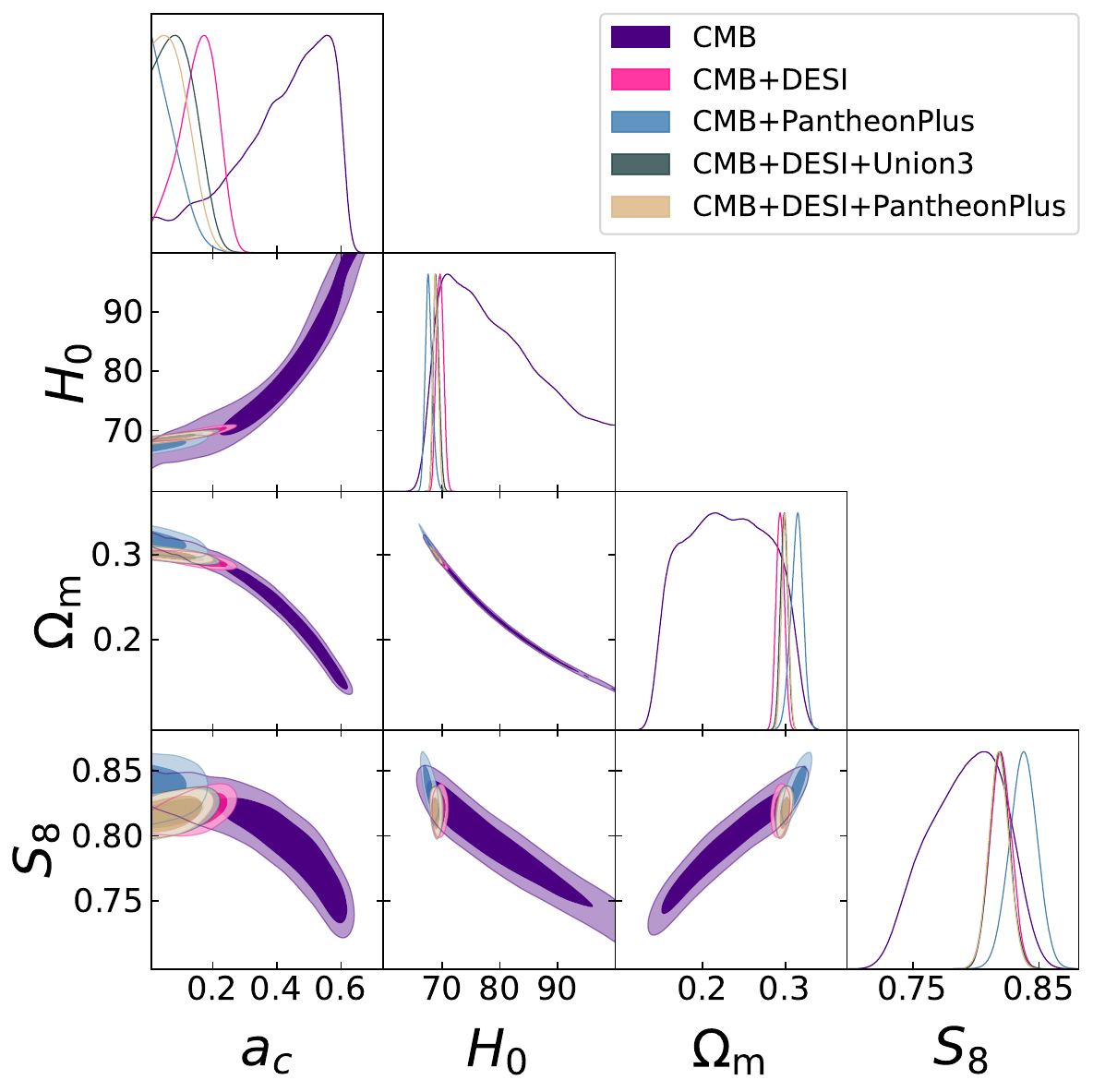}
    \caption{One-dimensional marginalized posterior distributions and two-dimensional joint confidence contours for selected CEDE model parameters, obtained using the combined CMB, DESI BAO, PantheonPlus, and Union3 datasets.}
    \label{fig:contour2}
\end{figure}

\section{Results}
\label{sec-results}

\begin{figure}
    \centering
    \includegraphics[width=1.0\linewidth]{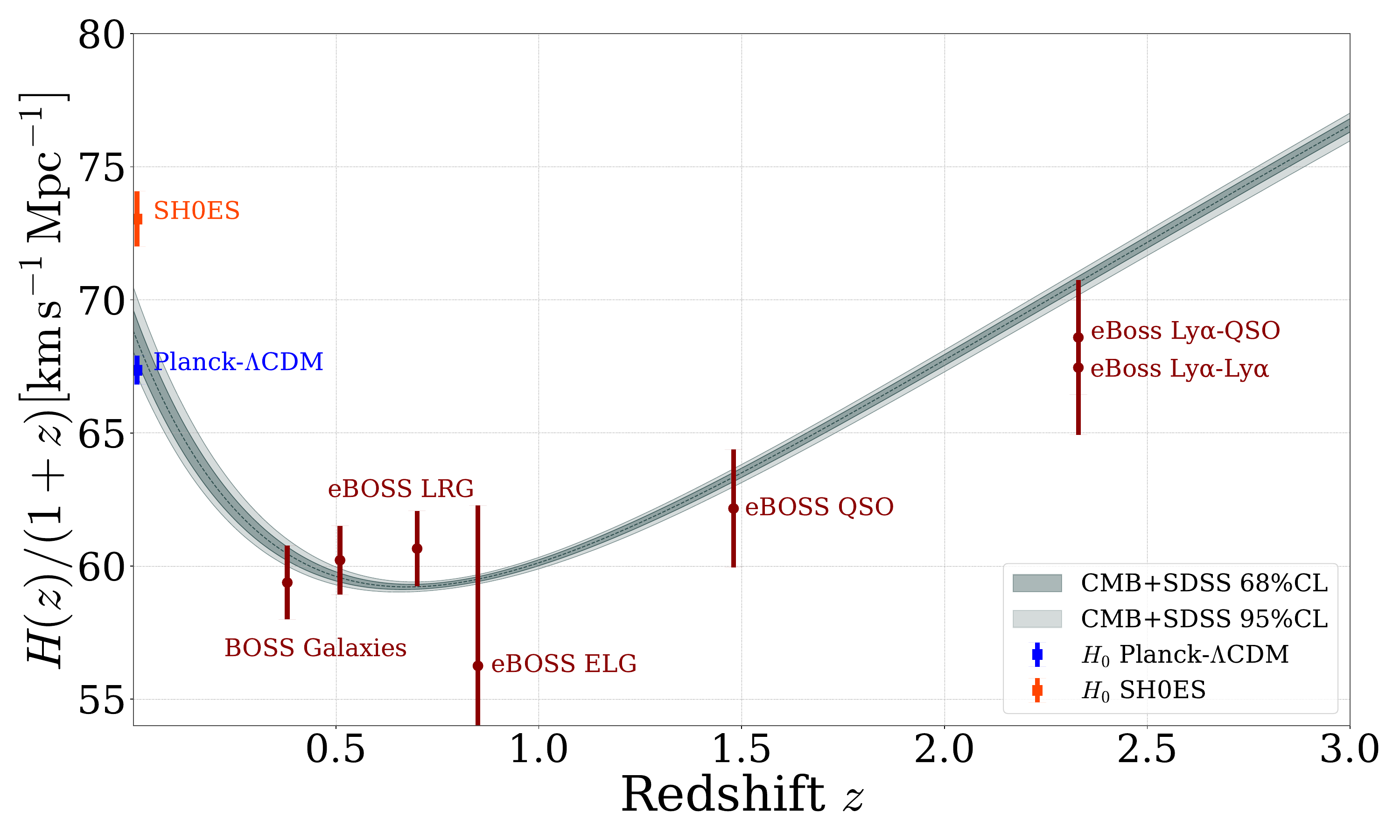}
\includegraphics[width=1.0\linewidth]{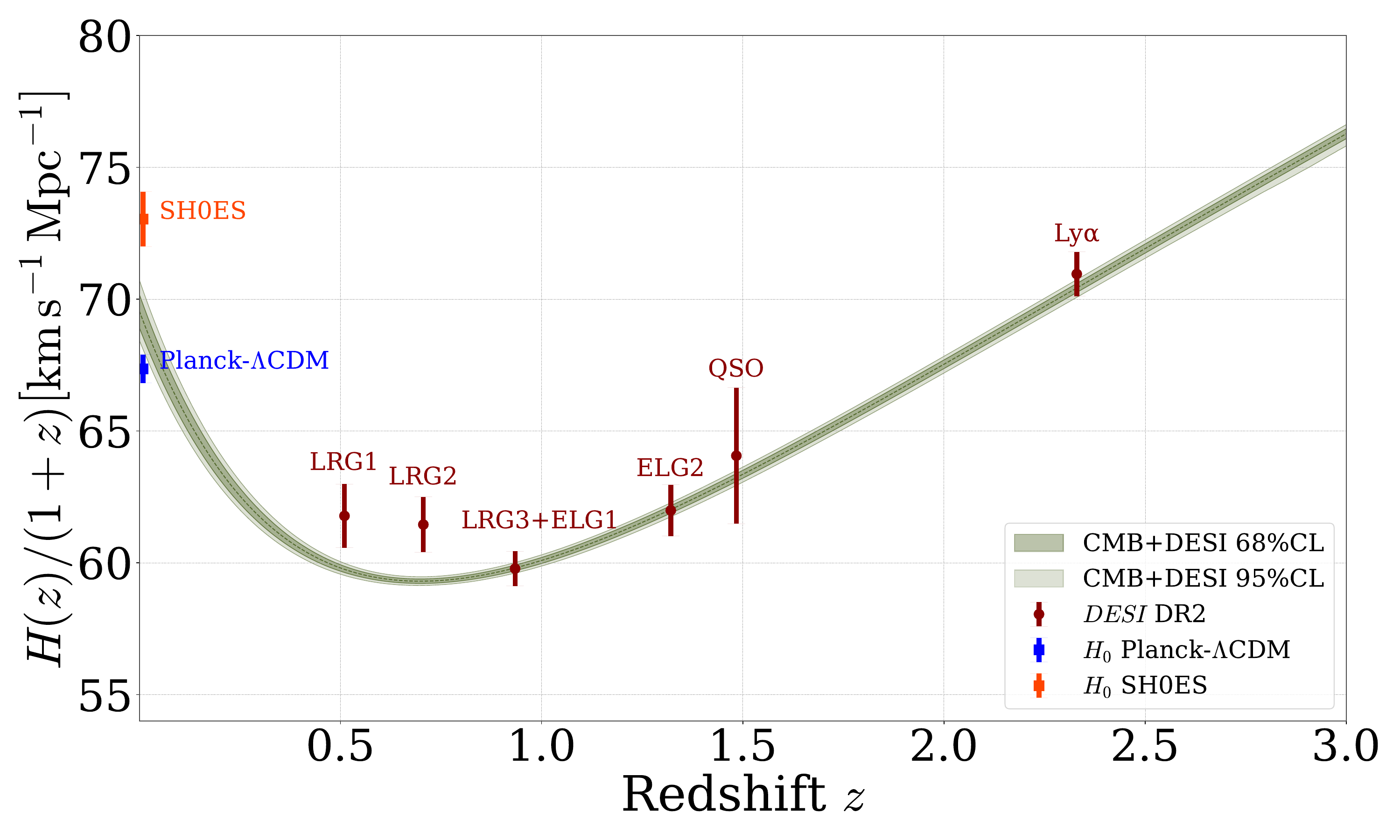}
    \caption{Rescaled Hubble parameter $H(z)/(1+z)$ showing the 68\% and 95\% confidence regions for the CEDE model obtained from CMB+SDSS (upper panel) and CMB+DESI (lower panel), compared with the corresponding SDSS and DESI BAO measurements~\cite{eBOSS_DR16_cosmo, DESI:2025zgx}.}
    \label{fig:Hz}
\end{figure}

\begin{figure*}
    \centering
    \includegraphics[width=0.85\linewidth]{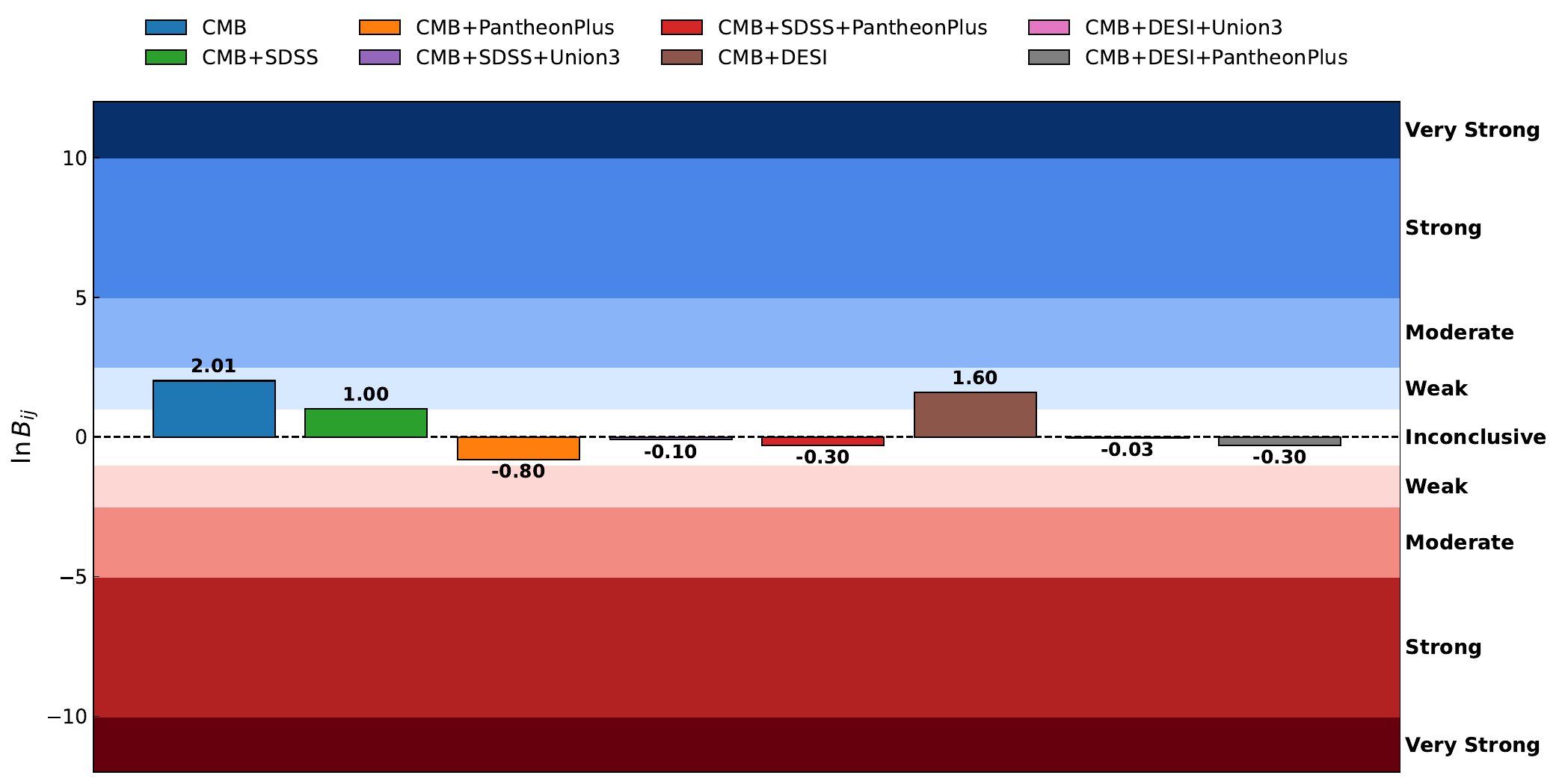}
    \caption{Values of $\ln B_{ij}$ ($i=$ CEDE, $j=$ $\Lambda$CDM) for the different observational datasets. Negative values of $\ln B_{ij}$ indicate a preference for the $\Lambda$CDM model, while $\ln B_{ij} > 0$ favors the CEDE scenario.} 
    \label{fig:BE}
\end{figure*}

We now present the constraints on the free and derived parameters of the CEDE scenario obtained using various cosmological datasets. Compared to the six-parameter $\Lambda$CDM model, CEDE introduces one additional parameter, $a_c$, which denotes the scale factor at which the dark energy phase transition occurs. Our analysis includes CMB data from \emph{Planck}~2018, both alone and in combination with baryon acoustic oscillation measurements from SDSS and DESI Data Release~2, as well as two independent Type~Ia supernova compilations, PantheonPlus and Union3.
In total, we perform eight distinct analyses for the CEDE model. The results are summarized in Tables~\ref{table:cons-1} and~\ref{table:cons-2}. Table~\ref{table:cons-1} reports constraints derived from CMB data alone and from its combination with SDSS BAO and the two SNIa compilations. Table~\ref{table:cons-2} presents the corresponding results obtained using CMB+PantheonPlus and those with DESI BAO in combination with CMB and the same SNIa datasets. The two tables therefore allow a direct comparison between SDSS- and DESI-based constraints.
The corresponding marginalized posterior contours are shown in Figs.~\ref{fig:contour1} and~\ref{fig:contour2}. In Fig.~\ref{fig:Hz}, we show the evolution of the rescaled Hubble parameter $H(z)/(1+z)$ for the CMB+SDSS and CMB+DESI combinations, compared with the corresponding SDSS and DESI measurements. Finally, Fig.~\ref{fig:BE} presents the Bayesian evidence results, illustrating the relative preference of the CEDE model with respect to the standard $\Lambda$CDM scenario.
In the following, we discuss in detail the constraints obtained from the different dataset combinations.

We begin by discussing the constraints obtained from CMB data alone (see Table~\ref{table:cons-1}). The CMB data favor the presence of a late-time phase transition in DE at several standard deviations, with a critical redshift $z_c \sim 1.4$, supporting the emergent nature of the CEDE scenario. Within this framework, the CMB-only analysis yields a high value of the Hubble constant, $H_0 = 80.2^{+5.1}_{-12.0}\,\mathrm{km\,s^{-1}\,Mpc^{-1}}$ at 68\% CL. Although the uncertainty is large, this value is consistent with direct late-time determinations, such as those reported by the H0DN, which finds $H_0 = 73.50 \pm 0.81\,\mathrm{km\,s^{-1}\,Mpc^{-1}}$ at 68\% CL~\cite{H0DN:2025lyy}.
This behavior is expected, since the CEDE model exhibits a phantom equation of state throughout the redshift range where DE is active, leading to a faster late-time expansion compared to non-phantom scenarios. Owing to the well-known anti-correlation between $H_0$ and the matter density, the higher value of $H_0$ inferred from CMB data corresponds to a lower matter density parameter, $\Omega_m = 0.230 \pm 0.048$ at 68\% CL. We also find a mildly reduced value of the clustering amplitude, $S_8 = 0.793^{+0.035}_{-0.026}$ at 68\% CL, compared to the $\Lambda$CDM prediction $S_8 = 0.834 \pm 0.016$ inferred from Planck TT,TE,EE+lowE.
Finally, both the improvement in the goodness of fit, $\Delta\chi^2 = -5.0$, and the Bayesian evidence, $\ln B_{ij} = 2.01$, indicate a preference for the CEDE model over $\Lambda$CDM when using CMB data alone. According to the Jeffreys' scale, however, this preference remains weak, as illustrated in Fig.~\ref{fig:BE}.

Including SDSS BAO data in combination with the CMB significantly alters the inferred properties of the CEDE model. In particular, the critical scale factor is shifted to earlier times, with $a_c = 0.138^{+0.069}_{-0.060}$ at 68\% CL. The combined CMB+SDSS dataset still allows for the presence of a DE phase transition, with $a_c \neq 0$ at slightly more than the $2\sigma$ level; however, the transition is now inferred to occur at a much higher redshift, $z_c \sim 6.2$, in contrast to the late-time transition favored by CMB data alone.
At the same time, the inferred value of the Hubble constant is reduced, yielding $H_0 = 68.85^{+0.67}_{-0.87}\,\mathrm{km\,s^{-1}\,Mpc^{-1}}$ at 68\% CL, which is only mildly higher than the $\Lambda$CDM value inferred from Planck data~\cite{Planck:2018vyg}. The corresponding constraints on $\Omega_m$ and $S_8$ are found to be very similar to those obtained within the Planck--$\Lambda$CDM framework. From the perspective of model comparison, both $\Delta\chi^2$ and the Bayes factor indicate a preference for CEDE over $\Lambda$CDM; however, according to the Bayesian evidence, this preference remains weak.
In contrast, when PantheonPlus supernova data are combined with the CMB (CMB+PantheonPlus), only an upper bound on $a_c$ is obtained, and no statistically significant evidence for a DE phase transition is found. In this case, the inferred value of $H_0$ is fully consistent with the Planck $\Lambda$CDM result~\cite{Planck:2018vyg}, while the remaining cosmological parameters, including $\Omega_m$ and $S_8$, closely match their Planck-$\Lambda$CDM values. Consistently, both $\Delta\chi^2$ and $\ln B_{ij}$ indicate a preference for $\Lambda$CDM over CEDE; nevertheless, this preference is not decisive when assessed through Bayesian evidence alone. 

We now turn to the combined datasets CMB+SDSS+Union3 and CMB+SDSS+PantheonPlus. In both cases, only an upper bound on the transition scale factor $a_c$ is obtained, indicating no evidence in favor of a dark energy phase transition. This behavior closely mirrors the results found for the CMB+PantheonPlus combination discussed above.
The inferred values of the Hubble constant are only very mildly shifted relative to the Planck $\Lambda$CDM determination and remain consistent within $1\sigma$. Likewise, the constraints on other derived parameters, such as $\Omega_m$ and $S_8$, are essentially unchanged and closely match the Planck $\Lambda$CDM estimates~\cite{Planck:2018vyg}.
From the perspective of model comparison, the statistical indicators show that the CEDE model is effectively indistinguishable from $\Lambda$CDM for these dataset combinations.
 
We now consider the remaining set of analyses including DESI BAO data. In Table~\ref{table:cons-2}, we present the constraints on the CEDE model obtained from the CMB+DESI, CMB+DESI+Union3, and CMB+DESI+PantheonPlus combinations. For the CMB+DESI dataset, we find statistically significant evidence for a non-zero transition scale factor, with $a_c = 0.154^{+0.070}_{-0.048}$ at 68\% CL, corresponding to a transition redshift $z_c \sim 5.5$. Compared to the CMB+SDSS case, this indicates that the dark energy phase transition occurs at a later epoch than that inferred when SDSS BAO data are included ($z_c \sim 6.2$).
This behavior can be understood as a consequence of the fact that DESI BAO data favor a lower value of the matter density parameter $\Omega_m$ compared to SDSS. When combined with CMB data, this preference allows for a higher value of the Hubble constant, yielding $H_0 = 69.56 \pm 0.61\,\mathrm{km\,s^{-1}\,Mpc^{-1}}$ at 68\% CL. As a result, the Hubble tension is mildly reduced, though it remains significant at the level of $\sim 3.9\sigma$. The lower value of $\Omega_m$ preferred by DESI is also reflected in a correspondingly lower value of the clustering amplitude $S_8$.
The two-dimensional posterior distributions in Fig.~\ref{fig:contour2} show a positive correlation between $H_0$ and the transition scale factor $a_c$, such that higher values of the Hubble constant are associated with a later dark energy phase transition. This correlation explains why the DESI-driven preference for lower $\Omega_m$ translates into a larger value of $a_c$ relative to the SDSS-based analysis.
From the perspective of model comparison, both the improvement in the goodness of fit, $\Delta\chi^2 = -3.2$, and the Bayes factor, $\ln B_{ij} = 1.6$, indicate a preference for the CEDE model over $\Lambda$CDM for the CMB+DESI dataset. According to the Jeffreys' scale, however, this preference remains weak.

For the combined analysis including Union3 supernovae, namely CMB+DESI+Union3, we find a noticeable shift in the transition scale factor, with $a_c = 0.097^{+0.031}_{-0.080}$ at 68\% CL, corresponding to a transition redshift $z_c \sim 9.3$. This implies $a_c \neq 0$ at slightly more than the 68\% confidence level. This result is particularly informative when compared to the CMB+SDSS+Union3 case, for which $a_c$ remained unconstrained.
The difference does not reflect a generic increase in constraining power of DESI relative to SDSS, but rather arises from the interplay between the supernova data and the BAO measurements. Union3 supernovae favor relatively higher values of the matter density parameter $\Omega_m$, which are compatible with SDSS BAO constraints and therefore do not force a preference for a specific value of $a_c$. In contrast, DESI BAO data prefer lower values of $\Omega_m$, leading to a tension with the Union3-preferred region. Owing to the anti-correlation between $\Omega_m$ and $a_c$ in the CEDE model (see Figs.~\ref{fig:contour1} and~\ref{fig:contour2}), this tension is resolved by shifting the posterior toward smaller values of $a_c$, corresponding to an earlier dark energy phase transition. As a result, among all dataset combinations considered, CMB+DESI+Union3 favors the earliest transition epoch.
On the other hand, for the CMB+DESI+PantheonPlus combination, only an upper bound on $a_c$ is obtained, similarly to what was found for CMB+SDSS+PantheonPlus (see Table~\ref{table:cons-1}). The constraints on $H_0$, $\Omega_m$, and $S_8$ for both CMB+DESI+Union3 and CMB+DESI+PantheonPlus are broadly consistent with each other, with the inferred values of $H_0$ being slightly higher than the Planck 2018 $\Lambda$CDM estimate~\cite{Planck:2018vyg}. According to the statistical indicators based on $\Delta\chi^2$ and $\ln B_{ij}$, these dataset combinations do not allow a statistically significant distinction between the CEDE and $\Lambda$CDM models.

Finally, a direct comparison between the SDSS and DESI BAO datasets provides useful insight into the role of BAO measurements in constraining the CEDE model.
The inclusion of BAO data in combination with CMB observations is particularly valuable, as BAO measurements help to break the geometric degeneracies inherent in the CMB alone. As a result, the joint CMB+BAO datasets can provide significantly tighter and more informative constraints than those obtained from CMB data alone.
Exploring all relevant combinations considered in this work, using two distinct BAO compilations from SDSS and DESI DR2, we find that both CMB+SDSS and CMB+DESI show a preference for the CEDE model over the standard $\Lambda$CDM scenario (see Fig.~\ref{fig:BE}).\footnote{Although the CMB+DESI+Union3 combination yields $\Delta\chi^2 < 0$ and $\ln B_{ij} > 0$, indicating a formal preference for CEDE, the corresponding strength of evidence is weaker than for CMB+DESI and CMB+SDSS. For this reason, we do not place CMB+DESI+Union3 on the same footing as these two dataset combinations.} We therefore focus in this section on the CMB+SDSS and CMB+DESI datasets in order to further assess the performance of the CEDE model.
In Fig.~\ref{fig:Hz}, we present the reconstructed expansion history in terms of $H(z)/(1+z)$ for the CEDE model, using CMB+SDSS (upper panel) and CMB+DESI (lower panel), and compare the results with the corresponding SDSS and DESI BAO measurements. While both BAO datasets are broadly consistent with the CEDE predictions, noticeable differences emerge between the two cases. In particular, DESI provides a better fit in the high-redshift regime ($z \gtrsim 2$) compared to SDSS.
Overall, when considering the expansion history up to $z \simeq 3$, the DESI data lead to an improved agreement with the CEDE model relative to SDSS.

\begin{table*}
\begin{center}
\resizebox{0.9\textwidth}{!}{%
\begin{tabular}{c||c|c|c|c}
\hline\hline
\textbf{Parameters} & \textbf{CMB} & \textbf{CMB+SDSS} & \textbf{CMB+SDSS+Union3} & \textbf{CMB+SDSS+PantheonPlus} \\
\hline\hline
\textbf{\boldmath$\log(10^{10} A_\mathrm{s})$} & $3.037\pm 0.014$ & $3.044\pm 0.013$ & $3.046\pm 0.013$ & $3.046^{+0.012}_{-0.014}$ \\ 
\textbf{\boldmath$n_\mathrm{s}$} & $0.9668\pm 0.0042$ & $0.9651\pm 0.0039$ & $0.9652\pm 0.0036$ & $0.9653\pm 0.0038$ \\
\textbf{\boldmath$100\theta_\mathrm{MC}$} & $1.04100\pm 0.00030$ & $1.04090\pm 0.00030$ & $1.04091\pm 0.00030$ & $1.04092\pm 0.00030$ \\
\textbf{\boldmath$\Omega_\mathrm{b} h^2$} & $0.02243\pm 0.00015$ & $0.02238\pm 0.00014$ & $0.02237\pm 0.00013$ & $0.02238\pm 0.00014$ \\
\textbf{\boldmath$\Omega_\mathrm{c} h^2$} & $0.1194\pm 0.0012$ & $0.12003\pm 0.00098$ & $0.12007\pm 0.00092$ & $0.12006\pm 0.00095$ \\
\textbf{\boldmath$a_{c}$} & $0.409^{+0.20}_{-0.069}$ & $0.138^{+0.069}_{-0.060}$ & $<0.110$ & $<0.100$ \\
\textbf{\boldmath$\tau_\mathrm{reio}$} & $0.0523\pm 0.0073$ & $0.0542\pm 0.0073$ & $0.0550\pm 0.0071$ & $0.0550^{+0.0068}_{-0.0078}$ \\
\textbf{$H_0$} & $80.2^{+5.1}_{-12}$ & $68.85^{+0.67}_{-0.87}$ & $68.15^{+0.49}_{-0.65}$ & $68.07^{+0.45}_{-0.55}$ \\
\textbf{$\Omega_\mathrm{m}$} & $0.230\pm 0.048$ & $0.3019\pm 0.0072$ & $0.3081\pm 0.0061$ & $0.3088\pm 0.0056$ \\
\textbf{$\sigma_8$} & $0.919^{+0.044}_{-0.097}$ & $0.8269^{+0.0085}_{-0.010}$ & $0.8209^{+0.0063}_{-0.0083}$ & $0.8199^{+0.0065}_{-0.0076}$ \\
\textbf{$S_8$} & $0.793^{+0.035}_{-0.026}$ & $0.8294\pm 0.0096$ & $0.8319\pm 0.0093$ & $0.8319\pm 0.0094$ \\
\textbf{$r_\mathrm{drag}$} & $147.20\pm 0.27$ & $147.08\pm 0.23$ & $147.08\pm 0.22$ & $147.08\pm 0.23$ \\
\hline
\textbf{$\chi^2$} & $2780.9$ & $2805.6$ & $2835.8$ & $4455.8$ \\
\textbf{$\Delta\chi^2$} & $-5.0$ & $-3.9$ & $0.01$ & $0.5$ \\
\textbf{$\ln B_{ij}$} & $2.01$ & $1.0$ & $-0.1$ & $-0.3$ \\
\hline\hline
\end{tabular}}
\end{center}
\caption{Observational constraints on the free and derived parameters of the CEDE scenario at 68\% CL (upper limits correspond to 68\% CL), obtained from CMB and SDSS BAO data in combination with different SNIa compilations.}
\label{table:cons-1}
\end{table*}

\begin{table*}
\begin{center}
\resizebox{0.9\textwidth}{!}{%
\begin{tabular}{c||c|c|c|c}
\hline\hline
\textbf{Parameters} & \textbf{CMB+PantheonPlus} & \textbf{CMB+DESI} & \textbf{CMB+DESI+Union3} & \textbf{CMB+DESI+PantheonPlus} \\
\hline\hline
\textbf{\boldmath$\log(10^{10} A_\mathrm{s})$} & $3.042\pm 0.013$ & $3.048\pm 0.014$ & $3.052\pm 0.014$ & $3.052\pm 0.013$ \\ 
\textbf{\boldmath$n_\mathrm{s}$} & $0.9637\pm 0.0039$ & $0.9677\pm 0.0037$ & $0.9687\pm 0.0035$ & $0.9687\pm 0.0036$ \\
\textbf{\boldmath$100\theta_\mathrm{MC}$} & $1.04083\pm 0.00031$ & $1.04105\pm 0.00029$ & $1.04111\pm 0.00029$ & $1.04111\pm 0.00029$ \\
\textbf{\boldmath$\Omega_\mathrm{b} h^2$} & $0.02232\pm 0.00014$ & $0.02245\pm 0.00013$ & $0.02247\pm 0.00013$ & $0.02248\pm 0.00013$ \\
\textbf{\boldmath$\Omega_\mathrm{c} h^2$} & $0.1207\pm 0.0011$ & $0.11903\pm 0.00092$ & $0.11869^{+0.00075}_{-0.00087}$ & $0.11863^{+0.00070}_{-0.00083}$ \\
\textbf{\boldmath$a_{c}$} & $<0.0821$ & $0.154^{+0.070}_{-0.048}$ & $0.097^{+0.031}_{-0.080}$ & $<0.103$ \\
\textbf{\boldmath$\tau_\mathrm{reio}$} & $0.0527\pm 0.0071$ & $0.0568\pm 0.0074$ & $0.0589\pm 0.0074$ & $0.0585\pm 0.0074$ \\
\textbf{$H_0$} & $67.65^{+0.51}_{-0.70}$ & $69.56\pm 0.61$ & $68.94^{+0.42}_{-0.56}$ & $68.79^{+0.39}_{-0.49}$ \\
\textbf{$\Omega_\mathrm{m}$} & $0.3140^{+0.0079}_{-0.0068}$ & $0.2938\pm 0.0048$ & $0.2984\pm 0.0044$ & $0.2996\pm 0.0041$ \\
\textbf{$\sigma_8$} & $0.8187^{+0.0055}_{-0.0076}$ & $0.8282\pm 0.0099$ & $0.8210^{+0.0071}_{-0.0091}$ & $0.8188^{+0.0066}_{-0.0084}$ \\
\textbf{$S_8$} & $0.837\pm 0.011$ & $0.8195\pm 0.0085$ & $0.8189\pm 0.0082$ & $0.8182\pm 0.0081$ \\
\textbf{$r_\mathrm{drag}$} & $146.98\pm 0.25$ & $147.27^{+0.22}_{-0.20}$ & $147.34\pm 0.21$ & $147.35\pm 0.21$ \\
\hline
\textbf{$\chi^2$} & $4189.6$ & $2796.6$ & $2828.6$ & $4206.4$ \\
\textbf{$\Delta\chi^2$} & $2.0$ & $-3.2$ & $-0.5$ & $0.5$ \\
\textbf{$\ln B_{ij}$} & $-0.8$ & $1.6$ & $-0.03$ & $-0.3$ \\
\hline\hline
\end{tabular}}
\end{center}
\caption{Observational constraints on the free and derived parameters of the CEDE scenario at 68\% CL (upper limits correspond to 68\% CL), obtained from DESI BAO data in combination with CMB and different SNIa compilations.}
\label{table:cons-2}
\end{table*}

\section{Summary and Discussions}
\label{sec-summary}

The dark sector of the Universe has remained one of the central puzzles of modern cosmology, despite decades of increasingly precise astronomical observations. The standard $\Lambda$CDM model has achieved remarkable success in describing a wide range of cosmological data; however, it is now confronted with several significant challenges. Among these, the tension in the Hubble constant $H_0$ between early-Universe determinations based on the CMB (within the $\Lambda$CDM framework) and direct late-time measurements from the distance ladder stands out as one of the most persistent and compelling discrepancies. This tension has motivated extensive efforts to assess whether extensions or modifications of the standard cosmological paradigm are required.
In parallel, a broad class of alternative cosmological models has been proposed and tested against the growing body of observational data. More recently, BAO measurements from DESI have provided evidence suggesting that the dark energy sector may exhibit time dependence, challenging the interpretation of dark energy as a simple cosmological constant with strictly constant energy density. As the volume and precision of cosmological datasets continue to increase, systematic tests of both established and novel cosmological models have become an essential component of modern cosmology, offering a broader perspective on the physical mechanisms governing the evolution of the Universe.

Motivated by these considerations, in this work we have investigated a well-defined extension of the dark energy sector, namely the critically emergent dark energy (CEDE) model~\cite{Banihashemi:2020wtb}, in which dark energy is effectively absent in the early Universe and emerges only at late times. This scenario belongs to the broader class of emergent dark energy models~\cite{Li:2019yem,Pan:2019hac,Li:2020ybr,Yang:2021eud,Benaoum:2020qsi}, but is distinguished by the presence of a cosmological phase transition that governs the onset of dark energy.
We have constrained the CEDE model using a comprehensive set of cosmological observations, including CMB measurements from \emph{Planck} 2018, two independent BAO compilations from SDSS and DESI DR2, and two Type Ia supernova samples, PantheonPlus and Union3. The resulting constraints on the free and derived cosmological parameters are summarized in Tables~\ref{table:cons-1} and \ref{table:cons-2}, while the corresponding marginalized posterior distributions and reconstructed expansion histories are shown in Figs.~\ref{fig:contour1}, \ref{fig:contour2}, and \ref{fig:Hz}. 

Our results indicate that evidence for a dark energy phase transition emerges from several dataset combinations, including CMB, CMB+SDSS, CMB+DESI, and CMB+DESI+Union3, albeit at different levels of statistical significance. Importantly, the inferred epoch of the transition is not universal but depends sensitively on the dataset combination employed. Using CMB data alone, we find a very late transition, occurring at $z_c \sim 1.4$. When BAO data are included, the preferred transition redshift shifts to significantly earlier epochs, with $z_c \sim 6.2$ for CMB+SDSS and $z_c \sim 5.5$ for CMB+DESI. This behavior can be understood in terms of the different matter density preferences of the BAO datasets. Both SDSS and DESI BAO measurements favor higher values of $\Omega_m$ than those preferred by CMB data alone, with DESI exhibiting a stronger pull toward lower $\Omega_m$ relative to SDSS. Through the intrinsic anti-correlation between $\Omega_m$ and the transition scale factor $a_c$ in the CEDE model, these shifts in the preferred matter density translate into corresponding changes in the inferred epoch of the dark energy phase transition.
For the remaining dataset combinations, we find that only CMB+DESI+Union3 is able to constrain the transition redshift, while in all other cases only upper limits on $a_c$ are obtained. Even in the CMB+DESI+Union3 case, however, the statistical significance remains modest, with $a_c \neq 0$ at slightly more than the $1\sigma$ level.

Turning to the key derived parameters, namely $H_0$ and $\Omega_m$, we find that for several dataset combinations—most notably CMB+SDSS and all combinations including DESI—the inferred value of the Hubble constant is mildly increased relative to the Planck $\Lambda$CDM determination. The largest shift is obtained for CMB+DESI, yielding $H_0 = 69.56 \pm 0.61\,\mathrm{km\,s^{-1}\,Mpc^{-1}}$ at 68\% CL. This leads to a partial reduction of the Hubble tension, which remains at the level of $\sim 3.9\sigma$ when quantified with respect to the H0DN value~\cite{H0DN:2025lyy}.
This increase in $H_0$ can be traced back to the preference of BAO data—particularly DESI—for lower values of the matter density parameter $\Omega_m$ when combined with CMB observations. Through the well-known parameter degeneracies, a lower $\Omega_m$ allows for a higher present-day expansion rate, naturally shifting the inferred value of $H_0$ upward within the CEDE framework.

In light of the model comparison statistics, our results present a nuanced picture. For several dataset combinations, we find $\Delta\chi^2 < 0$ together with $\ln B_{ij} > 0$, indicating a statistical preference for the CEDE model over the standard $\Lambda$CDM scenario. This outcome is particularly noteworthy, as it shows that CEDE can provide an improved description of the data in specific observational contexts. For other dataset combinations, however, the opposite trend is observed, with both indicators favoring $\Lambda$CDM instead. Taken together, these results suggest that while CEDE is not uniformly preferred across all datasets, it exhibits features that merit further investigation, especially given that only a limited class of dark energy models display comparable behavior in current cosmological analyses.

Overall, the findings of this work indicate that the CEDE scenario constitutes an interesting and partially competitive alternative to the standard $\Lambda$CDM framework. The recurrent hints of a late-time dark energy phase transition emerging from multiple dataset combinations motivate further theoretical and observational scrutiny of the physical mechanisms that could trigger such a transition. Future high-precision cosmological observations from ongoing and forthcoming surveys will play a crucial role in determining whether emergent dark energy scenarios such as CEDE represent a genuine extension of the standard cosmological paradigm or merely an effective phenomenological description of the current data.

\section*{Acknowledgments}
The authors acknowledge Abdolali Banihashemi for his insightful discussion.
EDV is supported by a Royal Society Dorothy Hodgkin Research Fellowship.
SP acknowledges the financial support from the Department of Science and Technology (DST), Govt. of India under the Scheme   ``Fund for Improvement of S\&T Infrastructure (FIST)'' (File No. SR/FST/MS-I/2019/41).  W. Yang's work is supported by the National Natural Science Foundation of China under Grant Nos. 12547110 and 12175096.  The authors acknowledge the use of High-Performance Computing resources from the IT Services at the University of Sheffield.
This article is based upon work from COST Action CA21136 Addressing observational tensions in cosmology with systematics and fundamental physics (CosmoVerse) supported by COST (European Cooperation in Science and Technology).

\bibliographystyle{apsrev4-1}
\bibliography{biblio}
\end{document}